\documentclass[11pt]{article}
\usepackage{graphicx}

\def\caltech{Kellogg Radiation Laboratory 106-38\\
California Institute of Technology, Pasadena, Ca 91125, USA}
\def\support{\footnote{Work supported by the Division of Physics, \\
  Mathematics and Astronomy, Caltech }}

\def\Title#1{\begin{center} {\Large #1 } \end{center}}
\def\Author#1{\begin{center}{ \sc #1} \end{center}}
\def\Address#1{\begin{center}{ \it #1} \end{center}}

\newenvironment{Abstract}{\begin{quotation}  }{\end{quotation}}
\newenvironment{Presented}{\begin{quotation} \begin{center} 
             PRESENTED AT\end{center}\bigskip 
      \begin{center}\begin{large}}{\end{large}\end{center} \end{quotation}}

\def\beq{\begin{equation}}
\def\eeq#1{\label{#1}\end{equation}}
\def\eeqn{\end{equation}}

\def\beqa{\begin{eqnarray}}
\def\eeqa#1{\label{#1}\end{eqnarray}}
\def\eeqan{\end{eqnarray}}

\let\bar=\overbar

\def\Dslash{\not{\hbox{\kern-4pt $D$}}}
\def\dslash{\not{\hbox{\kern-2pt $\del$}}}

\def\msb{{\bar{\ssstyle M \kern -1pt S}}}

\begin{document}
\begin{titlepage}

\vfill
\Title{Evaluation of reactor neutrino flux: issues and uncertainties }
\vfill
\Author{ Petr Vogel\support}
\Address{\caltech}
\vfill
\begin{Abstract}
Evaluation of the reactor $\bar{\nu}_e$ flux and spectrum is an essential ingredient
of their application in the neutrino oscillation studies. Two anomalies, i.e.
discrepancies between the observed and expected count rates, are widely discussed
at the resent time.
The total rate is $\sim$ 6\% lower than the expectation at all distances $>$ 10 m from the reactor.
And there is a shoulder (often referred to as ``bump") at neutrino energies 5-7 MeV, not
predicted in the calculated spectrum. I review the ways the flux and spectrum is 
evaluated and concentrate on the error budget. I argue that far reaching conclusions based
on these anomalies should await a thorough understanding of the uncertainties of the spectrum,
and point out possible standard physics sources of the anomalies. 
\end{Abstract}
\vfill
\begin{Presented}
NuPhys2015,  Prospects in Neutrino Physics \\
London, Barbican, Dec. 17, 2015
\end{Presented}
\vfill
\end{titlepage}
\def\thefootnote{\fnsymbol{footnote}}
\setcounter{footnote}{0}

\section{Introduction}

Nuclear reactor are pure and powerful sources of low-energy electron antineutrinos that have been successfully used
in the study of neutrino oscillations. In the past neutrino flux and spectrum was determined in a large number of experiments
at distances 10-100 m from the reactor core (see the review \cite{RMP}). Recent experiments, with considerably larger
and more accurate detectors, Daya-Bay \cite{Daya1}, RENO \cite{Reno1}, and Double-Chooz \cite{Dchooz1} 
 at $\sim$ 1 km were not only
able to determine the mixing angle $\theta_{13}$, but were also able to observe and confirm the anomalies mentioned
in the abstract.

\begin{figure}[htb]
\begin{center}
\includegraphics[width=5.0in]{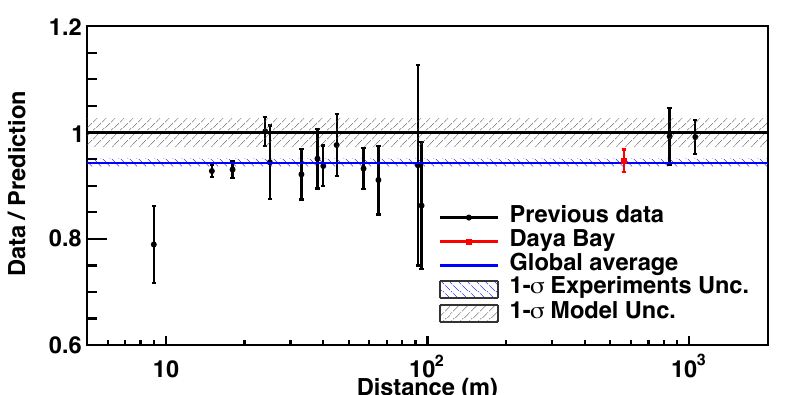}
\caption{The measured reactor rate, normalized to the prediction of \cite{mueller,huber} plotted
as a function of the distance from the reactor core. The rate is corrected for the 3-flavor
neutrino oscillation at each baseline. Reproduced from \cite{Daya1}.}
\label{fig:anomaly}
\end{center}
\end{figure}

The experimentally determined reactor antineutrino capture rate is shown in Fig. \ref{fig:anomaly} as a function of the
distance from the reactor. It is normalized to the prediction of refs. \cite{mueller,huber}; the global fit that includes all
past measurements and is corrected for the known three-flavor neutrino oscillations results in the ratio 
$R$ = 0.942 $\pm$ 0.009(exp) $\pm$ 0.025(model). Thus the average $R$ is experimentally determined with $\sim$1\% accuracy,
and suggests that 5-6\% of the $\bar{\nu}_e$ flux is missing already at $L \le$ 10 m.   Such disappearance at short distances
cannot be explained within the three-flavor mixing scenario with commonly accepted and experimentally well determined
parameters, and hence it is referred to as the ``reactor anomaly".

 If one would insist on explaining it using the neutrino oscillation phenomenon, it would require that the corresponding
 $L$(m)/$E_{\nu}$(MeV) is of the order of unity, i.e. it would imply
 the existence of one or more additional and necessarily sterile neutrinos with
 $\Delta m^2 \sim$ 1 eV$^2$. That hypothesis was raised, e.g., in \cite{mention}. In that sense the reactor anomaly is in accord
 with other indications of the $\nu_e$ non-standard disappearance (or mixing) such as LSND \cite{lsnd}, MiniBooNE \cite{miniboone},
 GALLEX \cite{gallex} or SAGE \cite{sage}. However, the significance of the reactor anomaly obviously critically depends on the
 value and the uncertainty of the predicted reactor $\bar{\nu}_e$ spectrum.
 
\begin{figure}[htb]
\begin{center}
\includegraphics[width=4.0in]{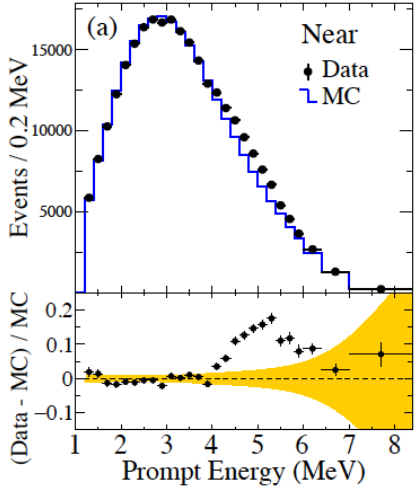}
\caption{The shoulder (bump) observed in the near detector  \cite{seo} at RENO.
The predictions are from the Huber-Mueller model \cite{huber, mueller} model, normalized to
the same number of events.
}
\label{fig:bump}
\end{center}
\end{figure}

In addition all three recent large reactor experiments, Daya-Bay, RENO, and Double Chooz  observed
a feature (or shoulder) in the 
experimental spectrum at 4-6 MeV of the prompt positron energy, $E_{prompt}\approx E_\nu+(M_p-M_n)+m_e$,
relative to the predicted theoretical evaluation in the refs. \cite{mueller,huber}.  
An example of the data, from the RENO experiment,
is shown in Fig. \ref{fig:bump}. The spectral shape of the shoulder cannot be produced by the
standard $L/E_{\nu}$ neutrino oscillations dependence; it is almost certainly caused by the
reactor fuel $\bar{\nu}_e$ emission. Note, however, that the shoulder was not observed in previous experiments.
We have to ask, therefore, whether its existence implies a 
significant problem of expected/predicted reactor spectra in general?

In the following I discuss first how the expected/predicted reactor spectrum is evaluated. That discussion then
leads to the consideration of the origin (or explanation) of these two observed anomalies. More detailed discussion
of these issues can be found in the forthcoming review \cite{annrev}.

\section{Nuclear reactors as $\bar{\nu}_e$ sources}

Nuclear reactors derive their power from the fission and 
from the radioactive decay of the corresponding fission fragments. 
In power reactors 99.9\% of the power comes from the fission
of $^{235}$U, $^{239}$Pu, $^{241}$Pu and $^{238}$U, while research reactors 
use usually highly enriched $^{235}$U as fuel which is then the only fission source.
The beta decay of the neutron rich fission fragments is the source of the electron antineutrinos.
The neutrino spectrum can be expressed as
\begin{equation}
S(E_{\nu}) = \frac{W_{th}}{\Sigma_i (f_i/F) e_i} \Sigma_i \frac{f_i}{F} \left( \frac {dN_i}{dE_{\nu}} \right) ~.
\label{eq:spectrum}
\end{equation}
Here $W_{th}$ is the reactor total thermal energy, $f_i/F$ is the fraction of fissions from actinide $i$,
$e_i$ is the effective thermal energy per fission, and $dN_i/dE_{\nu}$ is 
the cumulative $\bar{\nu}_e$ spectrum of $i$ normalized per fission. $W_{th}$ and $f_i/F$, which are
changing with time, are usually supplied by the reactor operators, and $e_i$ are known with negligible 
error. It is usually assumed that the individual neutrino spectra  $dN_i/dE_{\nu}$ depend only on the
nuclear properties of the isotope $i$. Their determination is the main topic of this section.

In the {\it `ab initio'} approach the aggregate fission antineutrino spectrum is determined by
summing the contributions of all $\beta$-decay branches of all fission
fragments
\begin{equation}
\frac{dN_i}{dE_{\bar{\nu}}} = \Sigma_n Y_n (Z,A,t) \Sigma_j b_{n,j}(E_0^j) 
P_{\bar{\nu}} (E_{\bar{\nu}}, E_0^j, Z) ~,
\label{fun1}
\end{equation}
where $Y_n (Z,A,t)$ is the number of $\beta$ decays of the fragment $Z,A$ at time $t$, 
and the label $n$ characterizes each fragment.
The quantity $Y_n$ converges to the cumulative fission yield.
The beta-decay branching ratios $b_{n,j}(E_0^j)$ are characterized by the endpoint energies $E_0^j$
and are normalized to unity, $\Sigma_j b_{n,j}(E_0^j) = 1$.
The function $P_{\bar{\nu}} (E_{\bar{\nu}}, E_0, Z)$ is the normalized 
$\bar{\nu}_e$ spectrum shape for the branch $n,j$. For the corresponding aggregate fission
electron spectrum, the $E_{\bar{\nu}}$ in the individual spectra $P$ must be replaced by
$E_e = E_0^j - E_{\bar{\nu}}$.

In applying the summation technique several sources of uncertainty arise.
The fission yields $Y_n$ for many important fragments
involve large  uncertainties. The branching ratios $b_{n,i}$ are also not known for all fragments, and nor are the
 quantum numbers (spins and parity) of all of the initial and final states.
The shape of the $\beta$ decay spectrum $P$ is simple for  Gamow-Teller allowed transitions ($\Delta I \le 1, 
\pi_i \pi_f = 1)$ transitions. However, $\sim30\%$ of the transitions making up the aggregate spectra are known 
to be so-called first forbidden transitions, $(\Delta I \le 2, \pi_i \pi_f = -1)$,
and involve nuclear structure dependent combinations of several operators.
Finally, there are important, albeit small,
corrections to the beta-decay spectra arising from radiative, nuclear finite size, and weak magnetism effects, 
and these can also depend on the details of the transitions.

The alternative method of determining the spectra $dN_i/dE_{\bar{\nu}}$  begins with the experimentally measured
aggregate {\it electron} spectrum associated with the fission of each individual actinide $i$. 
The electron spectrum for thermal neutron fission of $^{235}$U, $^{239}$Pu and $^{24 1}$Pu were measured at ILL, Grenoble, France
in 1980's \cite{Schr2}. $^{238}$U fissions only with fast neutrons; its electron spectrum was
measured much later  \cite{Haag} with larger error barrs.
These experimentally determined electron spectra are automatically summed over all 
fission fragments and the corresponding 
$\beta$-decay branches, so no information on the fission yields and branching ratios is needed. 
It is necessary, however, to convert them into the $\bar{\nu}_e$ spectra. 

To convert a measured aggregate electron spectrum into an antineutrino spectrum, the spectrum is binned over an  energy grid. 
The spectrum in each bin is fitted to a fictitious $\beta$ decay, and therefore 
the aggregate spectrum is described by a sum of virtual $\beta$-decay branches of assumed spectral shapes. 
The conversion to the antineutrino
spectrum is then simply accomplished by replacing the energy $E_e$ in each branch by $E_0 - E_{\bar{\nu}}$.
The procedure guaranties that the experimental electron spectrum is well reproduced.
However,
the converted $\bar{\nu}_e$ depends to some degree on the assumptions made about the 
spectrum shapes $P_i$, whether they correspond to allowed or forbidden transitions, their $Z$ dependence, and the form of the
corrections arising from nuclear finite size and weak magnetism.
To avoid
sizable systematic errors when converting the electron spectrum, it is necessary to use the
data bases and evaluate the dependence of the average nuclear charge Z on the endpoint
energy discussed, e.g. in \cite{huber,vogel07}

\section{Electron and $\bar{\nu}_e$ spectra of individual $\beta$ decays}

The $\beta$-decay spectrum shape of each $\beta$ branch can be expressed as
\begin{equation}
P_{\bar{\nu}} (E_{\bar{\nu}}, E_0^i, Z) =
K p_e E_e (E_0 - E_e)^2 F(Z,E_e) C(Z,E_e) (1 + \delta(Z,A,E_e)) ~,
\end{equation}
where $K$ is the normalization factor, 
$p_e E_e (E_0 - E_e)^2$ is the phase space factor, $F(Z, E_e)$ is the Fermi function that takes into account the
effect of the Coulomb field of the daughter nucleus on the outgoing electron, and the shape factor $C(Z,E_e)$ accounts for the energy
or momentum dependence of the nuclear matrix elements. For the allowed decays, $C(Z,E_e) = 1$. Finally, the
function $\delta(Z,A,E_e)$ describes the subdominant corrections to the spectrum shape.

For the allowed $\beta$ decays the corrections $ \delta(Z,A,E_e) = \delta_{QED} + \delta_{FS} + \delta_{WM}$ represent the
few \% deviations from the standard spectrum shape related to the radiative, finite nuclear size and weak magnetism. The latter two
depend to some extent on the details of each transition. In practice they are replaced by the estimate of their average values; thus
they must be assigned a sizable uncertainty. The evaluation in refs. \cite{mueller,huber} takes this source of error presumably correctly
into account. In fact, the treatment of these corrections in \cite{mueller,huber} is responsible for a substantial part of the revision of the
predicted flux compared with the more approximate way it was treated previously.

The treatment of the first forbidden $\beta$ decays represents even more significant source of uncertainty, which is difficult to quantify.
The shape factors corresponding to different operators are listed in ref. \cite{anna}. That reference also lists the corresponding weak
magnetism corrections that are rather different from those for the allowed decays. The corresponding finite size corrections have 
not been consistently evaluated as yet. The main difficulty, though, is that in most first forbidden decay, unlike the allowed ones, 
more that one operator contributes, and interference among their contributions is expected. 
Even though, as many textbooks claim, the shape of the first forbidden $\beta$ decays is similar to the shape of the allowed decays,
deviations at few \% level are expected and common. The fact that the quantum numbers of many short lived fission fragments are not
known makes the quantitative analysis even more difficult.

\section{The ``bump" in reactor spectrum} 

The shoulder or so-called ``bump" mentioned earlier was not observed in previous experiment. In particular, it is not present in
the experimental electron spectra \cite{Schr2}. The shoulder could have its 
origin in several effects that are not included, or not included accurately, in the reactor spectrum
predictions \cite{mueller,huber}. 
The contribution
of $^{238}$U, that is only weakly constrained by the observed electron spectrum 
might not be accurate.  The harder neutron spectrum in
power reactors may lead to
different fission fragment distributions than in the very thermal ILL reactor used for the electron fission spectra measurements.
Alternatively, the measured electron spectra themselves \cite{Schr2},
which represent the basis for the antineutrino evaluations \cite{mueller,huber}, might
be incorrect.   

Several possible origins of the bump have been identified and investigated by different authors \cite{dwyer, hayes1},
but it was generally concluded that, without further experimental investigation,
it is impossible to determine which, if any or several, of the explanations are correct. 
In the {\it ab initio} summation method the necessary input are the fission yields, and 
two standard fission-yields libraries, JEFF-3.1.1 and ENDF/B-VII.1
differ \cite{hayes1} significantly in the
predicted yields of several nuclei dominating the shoulder region.
When the problems in the ENDF/B-VII.1 library were corrected, 
the predictions of the two databases are considerably closer, 
and agree within 6\% at all energies. 
Most significantly, neither database (corrected ENDF or JEFF)
now predict a bump relative to the measured $^{235}$U aggregate electron fission spectrum.

At present, the two most likely sources of the bump seem to be $^{238}$U or the hardness of the neutron spectrum.
The $^{238}$U spectrum is considerably harder in energy than that of the other actinides, 
and the ENDF/B-VII.1 and JEFF-3.1.1 libraries predict a bump
relative to the $^{238}$U antineutrino spectrum of \cite{mueller} and \cite{Haag}.
Thus, without experiments designed to isolate the contributions from
each actinide to the shoulder, $^{238}$U cannot be ruled out as a significant source of the bump.
The effect of the hardness of reactor neutron spectrum on the
antineutrino spectrum has never been tested directly. The PWR reactors used by Daya
Bay, RENO and Double Chooz are harder in energy than
the thermal spectrum of the ILL reactor, and involve
considerably larger epithermal components. 

The existence of the ``bump" has little effect on the extraction of the neutrino oscillation
parameters from the reactor experiments and it could be entirely
uncorrelated with the ``reactor anomaly".
However, it raises the very serious question of how well the antineutrino spectra are known,
and suggests that estimated uncertainties at the 1-2\% level are too optimistic.

\section{Reactor anomaly}

The reactor anomaly, mentioned in the introduction,
is one of several experimental results that contradicts the 
standard three-flavor neutrino oscillation paradigm.
Clearly, this is an issue of fundamental importance,
potentially  a source of the long sought after ``physics beyond the standard model". 
In the case of the reactor $\bar{\nu}_e$ capture rate the experimental data are quite firm, but the expectations
depend on an assumed reactor spectrum,  involving uncertainties that are difficult to determine reliably.

The summation method requires knowledge of both the decay spectra and fission yields for all 
of the fragments determining the spectra, and both inputs involve uncertainties.
For the decay of individual nuclei, the databases are incomplete because about 5\% of the nuclei are 
sufficiently far from the line of stability that no measurements of the spectra are available, and thus
modeling is necessary. The spectrum shape corrections involve sizable uncertainties. For the weak magnetism
correction they are typically estimated to be $\sim$ 20\% for the allowed decays and perhaps 30\% for
the forbidden ones. The finite size corrections  involve both the weak transition density $\rho_W$ and 
the charge density $\rho_{ch}$. Several density or radius approximations 
have been made in the literature and these differ from one another by about  50\%, the tentative uncertainty.
For forbidden decays that correction is even more uncertain.
The database fission yields are also
uncertain for many important nuclei. While it is difficult to estimate the uncertainty in the database fission yields, 
the tentative place uncertainty arising from their contribution to the summation method is $\sim 10\%$.

Though many of the uncertainties also apply when converting a measured electron spectrum to an antineutrino spectrum,
the situation is somewhat different since the fit must reproduce the electron spectrum. Nevertheless, when in ref. \cite{hayes1} 
different assumptions were made about which weak magnetism and shape factors should be applied 
to the non-uniquely forbidden component of the spectrum and with fits to the electron spectrum of equal statistical accuracy, 
the antineutrino spectrum was found to vary by as much as 4\%.
To determine the full effect of the uncertainties that apply to a conversion from a measured electron spectrum to an antineutrino spectrum,  
requires a detailed multi-parameter sensitivity 
study. In the absence of such a study, 5\% uncertainty on
the conversion method is tentatively assigned.

In general, conversion of measured aggregate electron fission spectra provide more accurate determinations of the
antineutrino spectra than do predictions from the databases. 
The database calculations do, however, provide means of estimating the relative importance
of theoretical corrections to the spectra and their uncertainties. 
Improving on the theoretical inputs to the spectra will be challenging. Thus, there is a clear need for new experiments.
Ideally, these should involve more than one reactor design and fuel enrichment, because the remaining issues will require
a better understanding of the role of the hardness of the reactor neutron spectrum and of  the four individual
actinides that make up total spectra.  
For the bump energy region, better measurements  of the $^{238}$U spectrum would be particularly valuable.   

This brief discussion suggests that it is possible, perhaps even likely, that both anomalies have their origin in standard physics.
Decisive experiments are planned, and some of them will have results soon. Until then a diligent work on reducing the
systematic uncertainties is clearly indicated.


\begin{thebibliography}{99}


\bibitem{RMP} 
C. Bemporad, G. Gratta and P. Vogel,
Rev.Mod. Phys. {\bf 74}, 297 (2002).

\bibitem{Daya1} 
F. P. An {\it et al.}, Phys. Rev. Lett. {\bf 108},171803 (2012); ibid {\bf 116}, 061801 (2016).

\bibitem{Reno1}
J. K. Ahn {\it et al.}, Phys. Rev. Lett {\bf 108}, 191802 (2012).;
Seon-See Seo, arXiv:1410:7987 (2014).

\bibitem{Dchooz1}
Y. Abe {\it et al.}, Phys. Lett. {\bf B735}, 51 (2014); JHEP 10 (2014) 086 [Erratum ibid. 02 (2015) 074].
 
\bibitem{mueller} T. A. Mueller {\it et al.},  Phys. Rev. C {\bf 83}, 054615 (2011).

\bibitem{huber} P. Huber, Phys. Rev. C {\bf 84}, 024617 (2011); erratum ibid. {\bf 85},029901 (2012).

\bibitem{mention} G. Mention {\it et al.}, Phys. Rev. D  {\bf 83},073006 (2011).

\bibitem{lsnd} A. Aguilar  {\it et al.}, Phys. Rev. D {\bf 64},112007 (2001).

\bibitem{miniboone} A. Aguilar {\it et al.}, Phys.Rev.Lett. {\bf 110},161801 (2013).

\bibitem{gallex} P. Anselmann P. {\it et al.},  Phys. Lett. {\bf B342},440 (1995);
 and   Kaether F. {\it et al.}, Phys.  Lett. {\bf B685},47 (2010).

\bibitem{sage}  J. N. Abdurashitov {\it et al.}, Phys. Rev. Lett. {\bf 77},4708 (1996); 
 and, Phys. Rev. C  {\bf 80},015807 (2009).

\bibitem{seo} S-H. Seo,{\it Neutrino 2014, AIP Conf. Proc., E. Kearns and G. Feldman ed.} p. 080002 (2015), and
J.H. Choi, {\it et al.}, arXiv:1511.05849v2 (2015).  

\bibitem{annrev} A. C. Hayes and P. Vogel, {\it submitted to Vol. 66 of the Annual Review of Nuclear and Particle Physics}.

\bibitem{Schr2} F. von Feilitzsch, A. A. Hahn and K.  Schreckenbach, 
 Phys. Lett {\bf 118B},365 (1982); K. Schreckenach {\it et al.}, Phys. Lett {\bf 160B},325 (1985); 
 A. A. Hahn {\it et al.}, Phys. Lett {\bf 218B},365 (1989).

\bibitem{Haag} N. Haag {\it et al.}, Phys. Rev. Lett {\bf 112},122501 (2014).

\bibitem{vogel07} P. Vogel, Phys. Rev. C {\bf 76},025504 (2007).

\bibitem{anna} A. C. Hayes {\it et al.}, Phys. Rev. Lett. {\bf 112}, 202501 (2014).

\bibitem{dwyer} D. A. Dwyer  and T. J.  Langford, Phys. Rev. Lett. {\bf 114},012502 (2015).

\bibitem{hayes1} A. C. Hayes  et al., Phys. Rev. D {\bf 92},033015 (2015).


\end{thebibliography}
\end{document}